# Receiver Design for OWC Orbital Angular Momentum Communication in Data Center Applications

Judy Kupferman and Shlomi Arnon
Electrical and Computer Engineering Department
Ben-Gurion University of the Negev
Beer Sheva, Israel
shlomi@bgu.ac.il

A data center (DC) is one of the main building blocks of modern information technology. Any intra DC network should strive to achieve high throughput, while continuing to be flexible in responding to changing traffic patterns. Optical wireless communication links on top of the fiber/wire infrastructure could offer flexible and dynamic reconfiguration of the network. In this paper we present the design of an optical wireless communication orbital angular momentum (OAM) receiver with detector array. The proposed receiver is described in detail, a numerical example is given, and alternate options for its construction are explored. For a given set of parameters we found that a three ring detector with optimized ring area is comparable to seven ring evenly spaced detector.

## I. INTRODUCTION

The internet, social media, cloud computing big data processing, and internet of things (IoT) and vehicles (IoV) as well as wireless watches, smartphones and tablets with applications such as WhatsApp, Instagram, Snapchat, WeChat, Imo, and of course Google Facebook and other social media have revolutionized society. The basic technology infrastructure of these applications is the data centers (DC): a centralized warehouse, either physical or virtual, for the storage, processing, analytics and dissemination of data and information. [1] However, a new generation of applications is emerging, which will require ever faster, flexible intra DC networks. Optical wireless communication [1-5] can afford enormous flexibility for various DC traffic patterns [6-8], substantially reduced power consumption, high bandwidth, high data-rates and low latency [4].

In addition, OWC can be used to enhance the performance of virtualization techniques, which allow load balancing [7,8]. Figure 1 depicts the scenario under consideration. In our scenario the conventional rack is augmented by an optical wireless communication (OWC) transceiver.

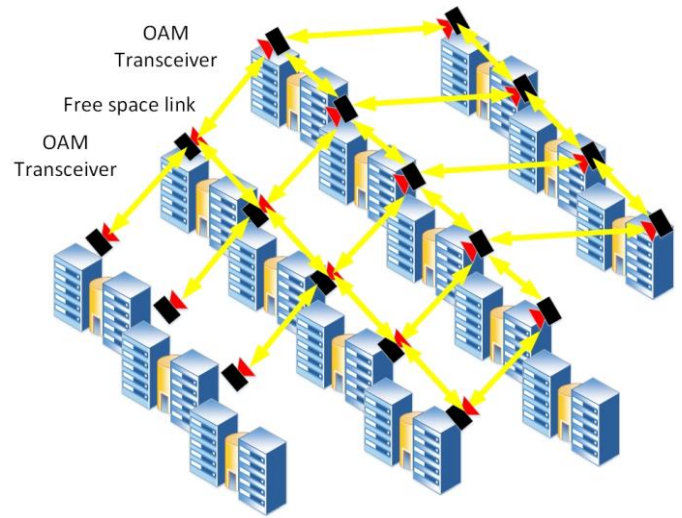

.

Fig. 1. Optical wireless OAM transciever enhances the intra data center network

The OWC transceiver can be installed on top of each rack and can communicate with neighboring transceivers to create an adaptive optical wireless network [4]. Creation of an optical link between the racks enables live migration. For improved flexibility, we propose a hybrid system that includes both free space and fiber. At the physical layer this involves point to point communication, with the capability to steer the transceiver to one of the 4 or 6 neighboring racks, This could make the FSO network flexible and adaptive dynamically to change in the traffic load. It is clear that the traffic between any two racks in the network could reroute at a higher layer. We assume that the average traffic rate will use fiber network but the peak traffic

---

will managed by free space optics. This gives the system maximal flexibility, in the ability to switch between the different options. We propose to use orbital angular momentum (OAM) for optical wireless communication. OAM multiplexing can access a theoretically unlimited set of OAM modes, and thus offer a much larger number of bits per symbol and a much faster data rate. Due to the controlled environment of the data center the distortion effect of the OAM could be predicted and mitigated.

The use of unique optical techniques in data centers is the focus of increasing interest [4,6,7,8,9]. Refs [10,11,12] describe OAM communication systems. A comprehensive review appears in [10] outlining the application of OAM to communications: beam generation, multiplexing, transmission through free space and fiber, networking and use in quantum information.

The rest of this paper is organized as follows. In Section II, we give a brief review of orbital angular momentum and then present details of the proposed communications scheme. In Section III we outline the proposed receiver design. Section IV describes a simulation of performance and error, followed by discussion and conclusions in Section V. Part of this work was presented at CNDSP2016 in Prague [5].

## II. OAM COMMUNICATION SYSTEM

Light beams can and generally do have orbital angular momentum (OAM), as shown in 1992 by Allen *et al*. [13] . The OAM of a light beam can be decomposed into a complete set of orthonormal Laguerre-Gauss modes which solve the paraxial Helmholtz equation in cylindrical coordinates. OAM modes can be generated in a variety of ways including holograms, cylindrical lens converters, phase plates, and SLMs [10]. A dedicated mechanism for decoding both intensity and phase has been developed [14,15]. Our system detects intensity distribution alone but is extremely simple. In the proposed system, an OWC transceiver is installed on top of each rack and can communicate with neighboring transceivers to create an adaptive optical wireless network [4]. Creation of an optical link between the racks allows the load to be rerouted. Fig.2 shows details of the scheme. Photons from the source go through a modulator, where they are brought to a specified state of OAM according to the modulation signal.

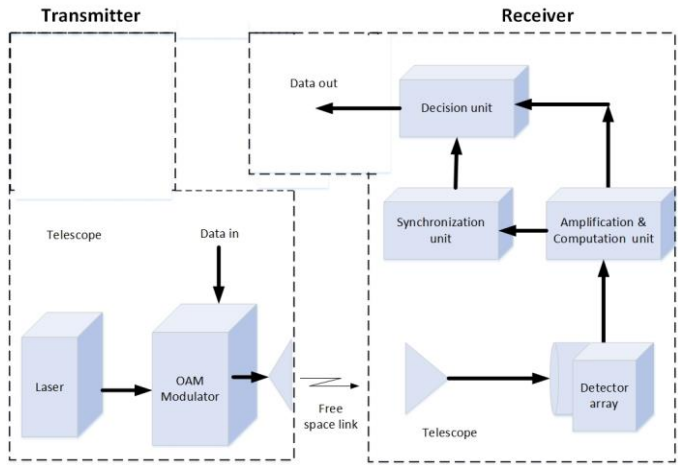

Fig. 2. The communication system scheme. Photons are transferred to a modulator where they are given OAM. According to the information input, the transmitter chooses one out of the *m* OAM modes. The beam then propagates through free space to the receiver unit, where it is directed to an array of detectors, converted to a digital signal, and analyzed to estimate the signal.

The modes are shown in Table1 along with the intensity distribution at various radii corresponding to those of the detector, and a graph of intensity distribution is shown in Fig.3. The intensities in the tables were computed by squaring the beam amplitude and integrating over ring area. Values are rounded off to second decimal. The images scale according to divergence of the beam. According to the information input, the transmitter chooses one out of *m* OAM modes. The beam then is collimated by a telescope and transmitted for a short distance in free space, to a telescope at the receiver unit. There the beam is directed to an array of annular detectors. The signal from the detectors then reaches a computation unit where it is digitized and analyzed, while the system is synchronized with the transmitter. The results of the analysis then pass through the decision unit where the correct modes are chosen and the information is extracted accordingly.

## III. DIGITAL DESIGN OF OAM RECEIVER DESIGN

In our proposed receiver design, we take advantage of the symmetry of OAM beam, thus our detector has radial symmetry. The detector is made up of PIN photodiodes arranged in a series of annular rings that fill the entire beam area (Fig. 4). The rings are not identical in radius, but rather they have been optimized for maximal distinction of the modes. Fig. 3 is a graph of intensity distribution for an alphabet of four OAM modes, showing the different intensity peaks at various distances from the center. From the figure one sees that these four modes have sufficiently distinct maxima to be suitable for the receiver, while graphs of other modes showed far too much overlap.

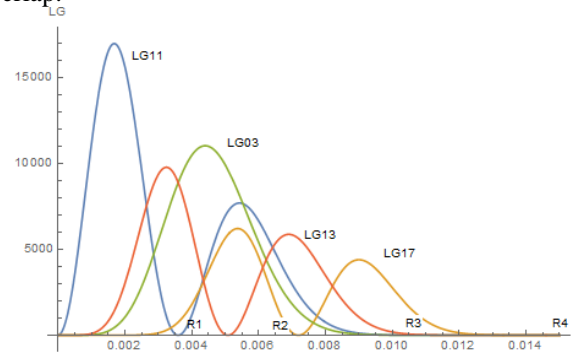

Fig. 3. Intensity distribution of the four OAM modes used in this scheme. R1,R2,R3 and R4 are the outer radii of the annular detector rings (Table I).

Table 1 gives the amplitude distribution for an alphabet of four modes and depicts the mapping of modes to symbols.

### A. Digital design of optimized three ring receiver

The signal from each ring enters the input of a transimpedance amplifier (TIA) for conversion of the sensor current to voltage. Then it reaches a comparator with a reference voltage. Each comparator will create logic "1" if the energy from its ring is above a certain value, in this case 0.3 of the total beam energy. The comparator output is sent to a set of simple logic gates consisting of NOTs, AND and OR, to extract the LSB and MSB (see Figure 4), where

$$LSB = C$$
$$MSB = \overline{B} + \overline{A}\overline{C}. \tag{1}$$

A,B and C refer to the signal from comparators from the first (inner), second, and third ring of detectors respectively. The radii were chosen in such a way that three are sufficient for mode identification. A fourth ring is used in order to detect error and misalignment. The two telescopes must be aligned, so that all the energy reaches the inner three rings. A signal from the fourth ring shows the beam is not properly aligned. The simplicity of logic design, using a very small number of logic gates, means that decoding of the signal is extremely fast and simple.

The calculation above was for radii that were optimized for clearest distinction between modes. Optimization was made for the four modes listed above and would not apply to any general set of modes. Clearly it would be simpler to manufacture a detector whose modes were evenly spaced. Therefore we also examined a setup with three rings that were evenly spaced rather than optimized, with the aim of comparing performance. However for the modes in question no threshold exists capable of distinguishing between modes.

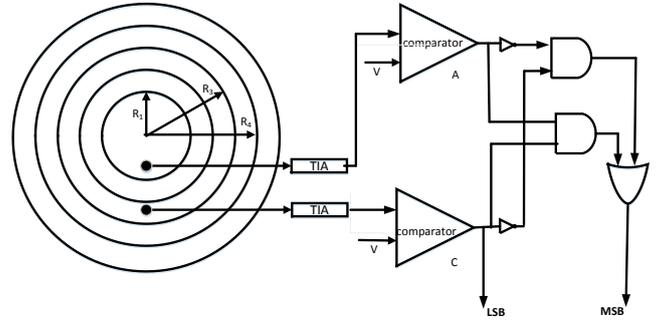

Fig. 5. Five ring OAM receiver scheme. The detector rings fill the entire beam area. The signal from each ring goes through a transimpediance amplifier (TIA), then to a comparator, and the LSB and MSB are extracted as in eq.(2).

### C. Digital design of seven ring receiver

In the case of a five ring evenly spaced detector, performance is not adequate. In hopes of improving this, we then attempted a detector with seven evenly spaced rings within the same total radius, to see if greater resolution would bring an advantage. The logic circuit in such a case is extremely simple:

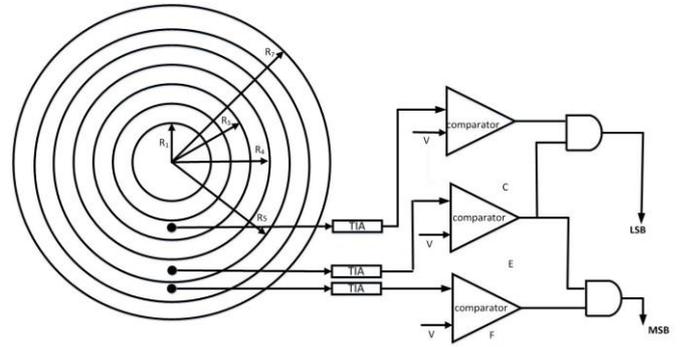

Fig. 6. Seven ring OAM receiver scheme. The detector rings fill the entire beam area. The signal from each ring goes through a transimpediance amplifier (TIA), then to a comparator, and the LSB and MSB are extracted as in eq.(2).

$$LSB = E + F$$
$$MSB = E + C. \quad (3)$$

A ten ring detector gives a somewhat simpler logic circuit:

$$LSB = \overline{D} + \overline{E}$$
$$MSB = C. \quad (4)$$

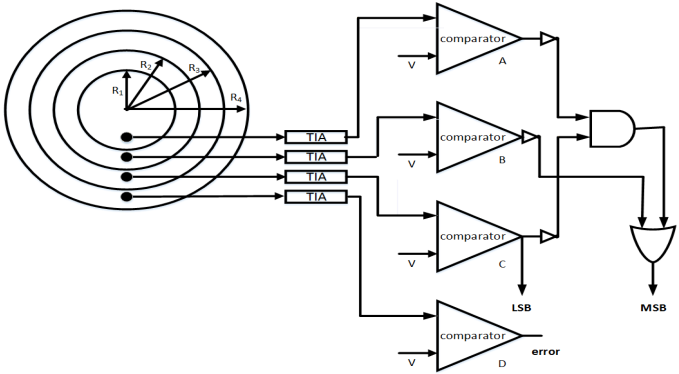

Fig. 4. The OAM receiver scheme. The detector is a series of concentric rings filling the entire beam area. The signal from each ring goes through a transimpediance amplifier (TIA), then to a comparator, and the LSB and MSB are extracted as in eq.(2).

### B. Digital design of five ring receiver

Another possibility is to construct a detector of the same size but with a larger number of evenly spaced rings rather than specifically optimized radii. Increasing the number of evenly spaced rings to five, within the same exterior radius, there exists a reasonable threshold of 0.2. This is because these particular four modes happen to differ sufficiently within four evenly spaced rings (the fifth is for calibration.) The logic circuits are

$$LSB = C$$
$$MSB = AC + \overline{A}\,\overline{C}. \quad (2)$$

In simulation of performance of detectors with different numbers of rings, it is necessary to take into account the different detector area in calculations of noise, because the variance is proportional to area. Therefore the noise was multiplied in all cases by the normalized ring area.

### IV. RESULTS

In the previous sections we presented a concept for a system that uses OAM in data centers. In order to consider its performance, we did a Monte Carlo simulation where we added additive white Gaussian noise (AWGN) to the signal and then calculated the symbol error probability. The symbol error rate was calculated as a function **α**, the ratio between the mean of the sum of signals from all detector rings and the root of the sum variance.

The threshold distinguishing logic values 1 and 0 was adapted optimally to the given **α**. Figures 7 and 8 show the MSB and LSB error probability as a function of **α**. The MSB

(LSB) error is 1/4 of the sum of the MSB (LSB) error for each mode. The receiver with optimized radii does not show better performance than that with five evenly spaced rings. Increasing the number of rings to seven brings improvement, and yet a ten ring receiver gives the best performance for MSB and the worst for LSB. This seems to be because performance is not just related to one factor, but to a combination: both to ring density and to the specific logic circuit for that detector.

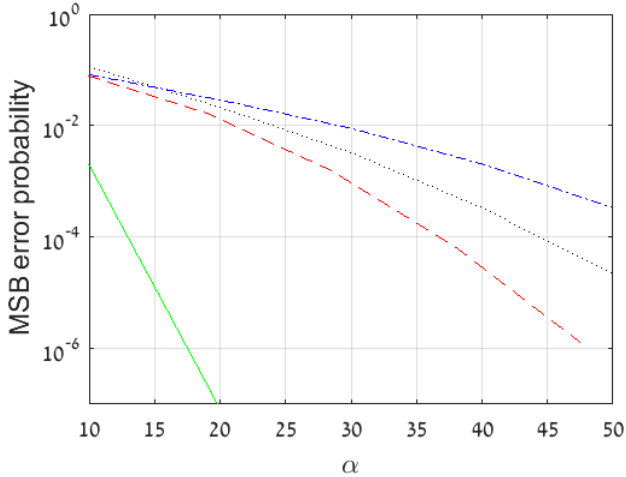

Fig. 7. Comparison of error in MSB as function of α the total normalized signal power for three optimized (black, dotted), 5 evenly spaced rings (blue dot-dash line) and 7 evenly spaced rings (red dashed line), and 10 evenly spaced rings (green solid line). The y axis is logarithmic.

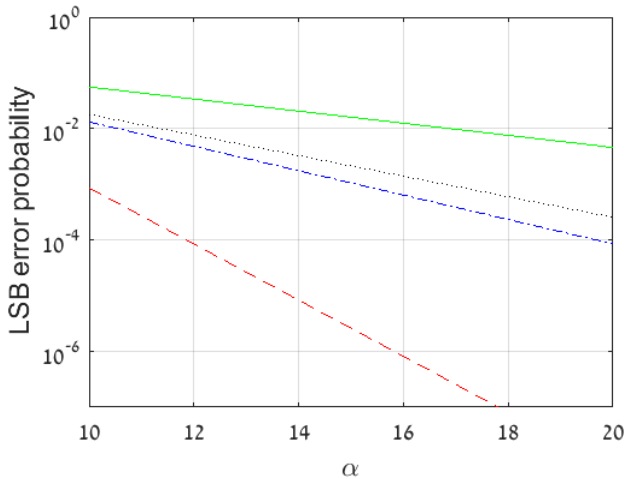

Fig. 8. . Comparison of error in LSB as function of α the total normalized signal power for three optimized rings (black, dotted), 5 evenly spaced rings (blue dot-dash line) and 7 evenly spaced rings (red dashed line), and 10 evenly spaced rings (green solid line). The y axis is logarithmic.

## V. SUMMARY AND CONCLUSIONS

We found that by increasing ring resolution from three to seven for the given setup, performance was improved for both MSB and LSB. In addition, it seems that optimization of radii can lead to moderate improvement in performance, which can also be achieved with a very small increase in ring number without necessitating special detectors for each set of modes. We also found that it is possible to design a very non complex logic system, even for a large number of rings.

In order to get higher accuracy as required for a data center, the probability for error should be on the order of $10^{-14}$-$10^{-18}$. However the trend in the numerical work was unchanged over a range of simulation iterations. These simulations were for four modes, however extension to a greater number of modes is straightforward.

We conclude that the value that was found for good SNR is not sufficiently low, due to overlap between modes. In order to reduce the required SNR the size of the dictionary could be reduced, since it would be much simpler to find two modes that are clearly distinguishable (for example in Fig.4 it is clear that LG11 and LG17 are easily distinct). Calculation of the probability of error for individual modes showed strong differences between them, depending on the distribution of intensity, so that restriction to fewer modes would lead to considerable improvement in performance. If the number of modes are reduced, additional degrees of freedom could be added using wavelength or polarization [13,15].

This is a general outline of concept. Implementation of the system requires a decrease in acceptable SNR, either by seeking further techniques for optimization, or increasing the number of detector rings, or by using a hybrid system with fewer OAM modes and additional degrees of freedom such as polarization or wavelength. In either case use of light for free space communication will increase flexibility and lower maintenance for the data center.

## *References*

[1] A.S.Hamza, J.S.Deogun, D.R.Alexander, "Wireless Communication in Data Centers: A Survey," IEEE Communications Surveys and Tutorials, V.18, 3 (2016)

[2] S. Arnon, M. Uysal, Z.Ghassemlooy, and J. Cheng. "Optical Wireless Communications," Selected Areas in Communications, IEEE Journal V 33, no. 9 (2015): 1733-1737.

[3] Z. Ghassemlooy, S. Arnon, M. Uysal, Z. Xu, and J. Cheng. "Emerging optical wireless communications-advances and challenges." Selected Areas in Communications, IEEE Journal V 33, no. 9 (2015): 1738-1749.

[4] S. Arnon. "Next-generation optical wireless communications for data centers." In SPIE OPTO, pp. 938703-938703. International Society for Optics and Photonics, 2015.

[5] J. Kupferman, Judy and S Arnon. "Receiver design for OWC orbital angular momentum communication in data center applications." In Communication Systems, Networks and Digital Signal Processing (CSNDSP), 2016 10th International Symposium on, pp. 1-6. IEEE, 2016.

[6] Z.Asad, M.A.R.Chaudhry, "A Two-Way Stret: Green Big Data Processing for a Greener Smart Grid," IEEE Systems Journal, V.PP, 99 (2016)

[7] J. Whitney and P. Delforge. "Data center efficiency assessment," Issue Paper on NRDC (the Natural Resource Defense Council), 2014.

[8] D. Halperin, S. Kandula, J. Padhye, P. Bahl, and D. Wetherall. "Augmenting data center networks with multi-gigabit wireless links," Presented at ACM SIGCOMM Computer Communication Review, 2011.

[9] I. Fujiwara, M. Koibuchi, T. Ozaki, H. Matsutani, and H. Casanova, "Augmenting low-latency HPC network with free-space optical links," Presented at High Performance Computer Architecture (HPCA), 2015 IEEE 21st International Symposium On. 2015, .

[10] Shlomi Arnon, and Moti Fridman. "Data center switch based on temporal cloaking." Lightwave Technology, Journal of 30, no. 21 (2012): 3427-3433.

[11] A.E. Willner, H. Huang, Y. Yan, Y. Ren, N. Ahmed, G. Xie, C. Bao, L. Li, Y. Cao, Z. Zhao, J. Wang, M.P.J. Lavery, M. Tur, S. Ramachandran, A. F. Molisch, N. Ashrafi, and S. Ashrafi, "Optical communications using orbital angular momentum beams," Adnvances in Optics and Photonics 7:1, 2015.

[12] Y. Yan, G. Xie, M.P.J. Lavery, H. Huang, N. Ahmed, C. Bao, Y. Ren, Y. Cao, L. Li, Z. Zhao, A.F. Molisch, M. Tur, M. J. Padgett, and A. E. Willner, "High-capacity millimetre-wave communications with orbital angular momentum multiplexing," Nature Communications 5, 4876, 2014.

[13] H. Huang, G. Xie, Y. Yan, N. Ahmed, Y. Ren, Y. Xue, D. Rogawski, M. J. Willner, B. I. Erkmen, K. M. Birnbaum, S. J. Dolinar, M.P.J. Lavery, M.J. Padgett, M. Tur, and A.E. Willner, "100 Tbit/s free-space data link enabled by three-dimensional multiplexing of orbital angular momentum, polarization, and wavelength," Opt.Lett. 39:2, 197, 2014.

[14] L. Allen, M. Beijersbergten, R. Spreeuw, and J. Woerdman, "Orbital angular momentum of light and the transformation of Laguerre-Gaussian laser modes," Phys. Rev. A 45, 81885-8189, 1992.

[15] D.L. Andrews and M. Babiker, eds., The Angular Momentum of Light, Cambridge University Press, 2013.

[16] M.P.J. Lavery, A. Fraine, D. Robertson, A. Sergienko, J. Courtial, A. E. Willner, and M. J. Padgett, "The measurement and generation of orbital angular momentum using an optical geometric transformation," Proc. SPIE 8610, Free-Space Laser Communication and Atmospheric Propagation XXV, 86100J, 2013.

[17] R. Fickler, R. Lapkiewicz, M. Huber, M.P.J. Lavery, M. J. Padgett, and A. Zeilinger, "Interface between path and orbital angular momentum entanglement for high-dimensional photonic quantum information," Nature Communications 5,4502, 2014.

[18] G.C.G. Berkhout, M.P.J. Lavery, J. Courtial, M.W. Beijersbergen, and M.J. Padgett, "Efficient Sorting of Orbital Angular Momentum States of Light," Phys. Rev. Lett 105, 15, 153601, 2010.

TABLE I.  ALPHABET OF FOUR OAM MODES, 3 RING DETECTOR

| Symbol | Logic values | | | Mode | Image | Normalized intensity | | |
|---|---|---|---|---|---|---|---|---|
| | Ring 1 (A) | Ring 2 (B) | Ring 3 (C) | | | Ring 1 (inner) | Ring 2 | Ring 3 |
| 00 | 1 | 1 | 0 | P=1,L=1 | 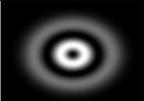 | 0.33 | 0.49 | 0.18 |
| 01 | 0 | 1 | 1 | P=1,L=7 | 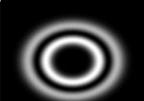 | 0.02 | 0.37 | 0.61 |
| 10 | 0 | 1 | 0 | P=0,L=3 | 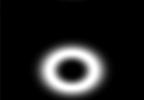 | 0.24 | 0.65 | 0.11 |
| 11 | 1 | 0 | 1 | P=1,L=3 | 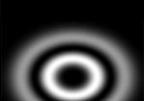 | 0.31 | 0.19 | 0.50 |